\documentstyle[aps,epsf]{revtex}

\begin{document}
\title{
  \begin{flushright} \begin{small}
    hep-th/0103240
  \end{small} \end{flushright}
Exact anisotropic brane cosmologies}
\author{Chiang-Mei Chen\footnote{E-mail: cmchen@phys.ntu.edu.tw}}
\address{Department of Physics, National Taiwan University,
         Taipei 106, Taiwan}
\author{T. Harko\footnote{E-mail: tcharko@hkusua.hku.hk}}
\address{Department of Physics, The University of Hong Kong,
         Pokfulam, Hong Kong}
\author{M. K. Mak\footnote{E-mail: mkmak@vtc.edu.hk}}
\address{Department of Physics, The Hong Kong University of Science
         and Technology, Clear Water Bay, Hong Kong}
\date{April 20, 2001}
\maketitle

\begin{abstract}
We present exact solutions of the gravitational field equations in
the generalized Randall-Sundrum model for an anisotropic brane
with Bianchi type I and V geometry, with perfect fluid and scalar
fields as matter sources. Under the assumption of a conformally
flat bulk (with vanishing Weyl tensor) for a cosmological fluid
obeying a linear barotropic equation of state the general solution
of the field equations can be expressed in an exact parametric
form for both Bianchi type I and V space-times. In the limiting
case of a stiff cosmological fluid with pressure equal to the
energy density, for a Bianchi type I Universe the solution of the
field equations are obtained in an exact analytic form. Several
classes of scalar field models evolution on the brane are also
considered, corresponding to different choices of the scalar field
potential. For all models the behavior of the observationally
important parameters like shear, anisotropy and deceleration
parameter is considered in detail.
\end{abstract}


\section{Introduction}
In two recent papers Randall and Sundrum \cite{RS99a,RS99b} have
shown that a scenario with an infinite fifth dimension in the
presence of a brane can generate a theory of gravity which mimics
purely four-dimensional gravity, both with respect to the
classical gravitational potential and with respect to
gravitational radiation. The gravitational self-couplings are not
significantly modified in this model. This result has been
obtained from the study of a single $3$-brane embedded in five
dimensions, with the $5D$ metric given by $ds^2 = e^{-f(y)}
\eta_{\mu\nu} dx^\mu dx^\nu + dy^2$, which can produce a large
hierarchy between the scale of particle physics and gravity due to
the appearance of the warp factor. Even if the fifth dimension is
uncompactified, standard $4D$ gravity is reproduced on the brane.
In contrast to the compactified case, this follows because the
near-brane geometry traps the massless graviton. Hence this model
allows the presence of large or even infinite non-compact extra
dimensions. Our brane is identified to a domain wall in a
$5$-dimensional anti-de Sitter space-time.

The Randall-Sundrum model was inspired by superstring theory. The
ten-dimensional $E_8 \times E_8$ heterotic string theory, which
contains the standard model of elementary particle, could be a
promising candidate for the description of the real Universe. This
theory is connected with an eleven-dimensional theory, M-theory,
compactified on the orbifold $R^{10} \times S^1/Z_2$ \cite{HW96}.
In this model we have two separated ten-dimensional manifolds.

The static Randall-Sundrum solution has been extended to
time-dependent solutions and their cosmological properties have
been extensively studied in papers too many to cite them all
\cite{KK00,BDEL00,BDL00,STW00,LMW00,KIS00,BDBL00,CEHS00} (for a
recent review of dynamics and geometry of brane Universes see
\cite{Ma01}). In one of the first cosmological applications of
this scenario it was pointed out that a model with a non-compact
fifth dimension is potentially viable, while the scenario which
might solve the hierarchy problem predicts a contracting Universe,
leading to a variety of cosmological problems \cite{CsGrKoTe99}.
By adding cosmological constants to the brane and bulk, the
problem of the correct behavior of the Hubble parameter on the
brane has been solved by Cline, Grojean and Servant
\cite{ClGrSe99}. As a result one also obtains normal expansion
during nucleosynthesis, but faster than normal expansion in the
very early Universe. The creation of a spherically symmetric
brane-world in AdS bulk has been considered, from a quantum
cosmological point of view, with the use of the Wheeler-de Witt
equation, by Anchordoqui, Nu\~{n}ez and Olsen \cite{AnNuOl00}.

The effective gravitational field equations on the brane world, in
which all the matter forces except gravity are confined on the
$3$-brane in a $5$-dimensional space-time with $Z_2$-symmetry have
been obtained, by using an elegant geometric approach, by
Shiromizu, Maeda and Sasaki \cite{SMS00,SSM00}. The correct
signature for gravity is provided by the brane with positive
tension. If the bulk space-time is exactly anti-de Sitter,
generically the matter on the brane is required to be spatially
homogeneous. The electric part of the $5$-dimensional Weyl tensor
$E_{IJ}$ gives the leading order corrections to the conventional
Einstein equations on the brane. The four-dimensional field
equations for the induced metric and scalar field on the
world-volume of a $3$-brane in the five-dimensional bulk with
Einstein gravity plus a self-interacting scalar field have been
derived by Maeda and Wands \cite{MW00}. The effective
four-dimensional Einstein equations include terms due to scalar
fields and gravitational waves in the bulk.

From a general theoretical point of view brane cosmological models
are two-measure theories, since they have two independent
Lagrangians, one associated to the brane, the other associated to
the bulk. An other two measure theory has been proposed by
Guendelman \cite{Gu199}, in which the energy-momentum tensor is
also a non-linear function of the general-relativistic
energy-momentum tensor. Due to the scale invariance, the true
vacuum state has zero energy density, when the theory is analyzed
in the conformal Einstein frame. Field theory models, constructed
in a gravitational theory where a measure of integration $\Phi$ in
the action is not necessary $\sqrt{-g}$, but is determined
dynamically through additional degrees of freedom, have been
presented in \cite{Gu299}. In this type of theory it is possible
to combine the solution of the cosmological constant problem with
the possibility of inflation and spontaneously broken gauge
unified theories. The models with global scale invariance allow
for non-trivial scalar field potentials and masses for particles,
so that the scale symmetry must be broken. The conserved
quantities have been obtained by Guendelman \cite{Gu100}, who
showed that the infrared behavior of the conserved currents is
singular, so there are no conserved charges associated with scale
symmetry (which implies that in some high field region the
potentials become flat). For closed strings and branes ( including
the supersymmetric case) the modified measure formulation is
possible and does not require the introduction of a particular
scale (the string or brane tension) from the beginning, but rather
these appears as integration constants \cite{Gu200}.

The linearized perturbation equations in the generalized
Randall-Sundrum model have been obtained, by using the covariant
nonlinear dynamical equations for the gravitational and matter
fields on the brane, by Maartens \cite{Ma00}. The nonlocal energy
density determines the tidal acceleration in the off-brane
direction and can oppose singularity formation via the generalized
Raychaudhuri equation. Isotropy of the cosmic microwave background
may no longer guarantee a Friedmann-Robertson-Walker geometry.
Vorticity on the brane decays as in general relativity, but
nonlocal bulk effects can source the gravitomagnetic field, so
that vector perturbations can also be generated in the absence of
vorticity.

The behavior of an anisotropic Bianchi type I brane-world in the
presence of inflationary scalar fields has been considered by
Maartens, Sahni and Saini \cite{MSS00}. According to their results
the magnitude of the anisotropy parameter of the brane does not
affect inflation, a large initial anisotropy introducing more
damping into the scalar field equation of motion. Inflation at
high energies proceeds at a higher rate than the corresponding
rate in general relativity. By numerical integration of field
equations it is also found that anisotropy always disappear within
a fixed interval of time, no matter what its initial value. The
shear dynamics in Bianchi type I cosmological model on a brane
with a perfect fluid has been studied in \cite{To01}. For
$1<\gamma <2$ the shear has a maximum at some moment during a
transition period from non-standard to standard cosmology, when
the matter energy density is comparable to the brane tension.

A systematic analysis, using dynamical systems techniques, of the
qualitative behavior of the Bianchi type I and V cosmological
models in the Randall-Sundrum brane world scenario, with matter on
the brane obeying a barotropic equation of state $p=(\gamma-1)
\rho$, has been performed by Campos and Sopuerta \cite{CS01}. In
particular, they constructed the state spaces for these models and
discussed what new critical points appear, the occurrence of
bifurcations and the dynamics of the anisotropy. Expanding Bianchi
type I and V brane-worlds always isotropize, although there could
be intermediate stages in which the anisotropy grows. Near the
Big-Bang the anisotropy dominates for $\gamma \leq 1$, in
opposition to the general relativistic situation when it dominates
for $\gamma<2$. Frolov \cite{F01} has shown that the geometrical
construction of the Randall and Sundrum brane world can be carried
out, without the assumption of spatial isotropy, by means of an
homogeneous and anisotropic Kasner type solution of the
Einstein-AdS equations in the bulk. The brane equations of motion
in this anisotropic space-time are solved by a static brane
configuration, with the geometry of the $3$-brane given by the
$4D$ Kasner solution. The set of sufficient conditions, which must
be satisfied by the brane matter and bulk metric so that a
homogeneous and anisotropic brane asymptotically evolves to a de
Sitter space-time in presence of a positive cosmological constant
have been derived in \cite{SaVeFe01}. In the presence of a
non-local energy density and/or a strong anisotropic stress (i.e.
a magnetic field), a Bianchi type I brane, even initially
expanding, is unstable and may collapse.

It is the purpose of this paper to investigate some classes of
exact solutions of the gravitational field equations in the brane
world model for the anisotropic Bianchi type I and V geometries,
for a conformally flat bulk (with vanishing Weyl tensor). For a
perfect cosmological fluid obeying a linear barotropic equation of
state, the general solution of the field equations can be obtained
in an exact parametric form, for arbitrary $\gamma$ or in an exact
analytic form for a stiff fluid. The inclusion of the quadratic
terms in the energy-momentum tensor of the perfect cosmological
fluids leads to major changes in the early dynamics of the
anisotropic Universe, as compared to the standard general
relativistic case. We also consider the dynamics of the scalar
fields on the brane for different scalar field potentials of
physical interest. The behavior of the observationally important
physical quantities like anisotropy, shear and deceleration
parameters is considered in detail for all these models.

The present paper is organized as follows. The field equations on
the brane are written down in Section II. In Section III we
present the general solution of the field equations for the
Bianchi type I geometry with a perfect cosmological fluid.
Evolution of the scalar fields in the brane-world is considered in
Section IV. Bianchi type V space-times are investigated in Section
V. In Section VI we discuss and conclude our results.

\section{Brane geometry, field equations and consequences}
In the $5D$ space-time the brane-world is located as $Y(X^I)=0$,
where $X^I, \, I=0,1,2,3,4$ are $5$-dimensional coordinates. The
effective action in five dimensions is \cite{MW00}
\begin{equation}
S = \int d^5X \sqrt{-g_5} \left( \frac1{2k_5^2} R_5 - \Lambda_5
\right) + \int_{Y=0} d^4x \sqrt{-g} \left( \frac1{k_5^2} K^\pm -
\lambda + L^{\text{matter}} \right),
\end{equation}
with $k_5^2=8\pi G_5$ the $5$-dimensional gravitational coupling
constant and where $x^\mu, \, \mu=0,1,2,3$ are the induced
$4$-dimensional brane world coordinates. $R_5$ is the $5D$
intrinsic curvature in the bulk and $K^\pm$ is the intrinsic
curvature on either side of the brane.

On the $5$-dimensional space-time (the bulk), with the negative
vacuum energy $\Lambda_5$ as only source of the gravitational
field the Einstein field equations are given by
\begin{equation}
G_{IJ} = k_5^2 T_{IJ}, \qquad T_{IJ}=-\Lambda_5 g_{IJ} + \delta(Y)
\left[ -\lambda g_{IJ} + T_{IJ}^{\text{matter}} \right],
\end{equation}
In this space-time a brane is a fixed point of the $Z_2$
symmetry. In the following capital Latin indices run in the range
$0,...,4$ while Greek indices take the values $0,...,3$.

Assuming a metric of the form $ds^2=(n_I n_J + g_{IJ})dx^I dx^J$,
with $n_I dx^I=d\chi$ the unit normal to the $\chi=\text{const.}$
hypersurfaces and $g_{IJ}$ the induced metric on
$\chi=\text{const.}$ hypersurfaces, the effective four-dimensional
gravitational equations on the brane take the form
\cite{SMS00,SSM00}:
\begin{equation}
G_{\mu\nu} = - \Lambda g_{\mu\nu} + k_4^2 T_{\mu\nu} + k_5^4
S_{\mu\nu} - E_{\mu\nu},
\end{equation}
where
\begin{equation}
S_{\mu\nu} = \frac1{12} T T_{\mu\nu} - \frac14 T_\mu{}^\alpha
T_{\nu\alpha} + \frac1{24} g_{\mu\nu} \left(
3T^{\alpha\beta}T_{\alpha\beta}-T^2 \right),
\end{equation}
and $\Lambda=k_5^2(\Lambda_5+k_5^2 \lambda^2/6)/2, k_4^2=k_5^4
\lambda/6$ and $E_{IJ}=C_{IAJB} n^A n^B$. $C_{IAJB}$ is the
$5$-dimensional Weyl tensor in the bulk and $\lambda$ is the
vacuum energy on the brane. $T_{\mu\nu}$ is the matter
energy-momentum tensor on the brane with components
$T^0{}_0=-\rho, T^1{}_1=T^2{}_2=T^3{}_3=p$ and $T=T^\mu{}_\mu$ is
the trace of the energy-momentum tensor.

In this paper we restrict our analysis to a conformally flat bulk
geometry with $C_{IAJB} \equiv 0$.

The Einstein equation in the bulk imply the conservation of the
energy momentum tensor of the matter on the brane,
\begin{equation}
T_\mu{}^\nu{}_{;\nu} \mid_{\chi=0} = 0.
\end{equation}

In the following we investigate the Bianchi type Universes that
are obtained from space-time geometries that admit a simple
transitive three-dimensional group of isometries $G_3$ on
space-like hypersurfaces. Hence the corresponding cosmological
models are spatially homogeneous. There is only one essential
dynamical coordinate, the time $t$, and the gravitational field
equations reduce to ordinary differential equations. the metric on
the brane is given in the general case by
\begin{equation}
ds^2 = - dt^2 + \gamma_{ab}(t) \, e^a_i(x) dx^i \, e^b_j(x) dx^j,
\end{equation}
where $e^a_i(x)$ are one-forms inverse to the spatial vector triad
$e_a^i(x)$, which have the same commutators $C^a{}_{bc}, \,
a,b,c=1,2,3$ as the structure constants of the group of isometries
and commute with the unit normal vector $e_0$ to the surfaces of
homogeneity, that is, $e_a = e_a^i \partial/\partial x^i$ and $e_0
= \partial/\partial t$ obey the commutation relations $[e_a, e_b]
= C^c{}_{ab} e_c, \, [e_0, e_a]=0$.

In order to simplify the calculations we only consider geometries
so that $\gamma_{ab}(t)$ can be taken to be a diagonal matrix. We
denote by $a_i(t), \, i=1,2,3$ the components of the metric
tensor on the anisotropic brane.

We define the following variable:
\begin{eqnarray}
V &=& \prod_{i=1}^3 a_i, \qquad \text{(volume scale factor),}
\label{V} \\
H_i &=& \frac{\dot a_i}{a_i}, \quad i=1,2,3, \qquad
\text{(directional Hubble parameters),} \label{Hi} \\
H &=& \frac13 \sum_{i=1}^3 H_i, \qquad \text{(mean Hubble
parameter),} \label{H} \\
\Delta H_i &=& H_i - H, \quad i=1,2,3.
\end{eqnarray}
From Eqs. (\ref{V}), (\ref{Hi}) and (\ref{H}) we obtain $H=\dot
V/3V$.

The physical quantities of observational interest in cosmology are
the expansion scalar $\theta$, the mean anisotropy parameter $A$,
the shear scalar $\sigma^2$ and the deceleration parameter $q$,
all are defined according to
\begin{eqnarray}
\theta &=& 3H, \\
A &=& \frac13 \sum_{i=1}^3 \left( \frac{\Delta H_i}{H} \right)^2,\\
\sigma^2 &=& \frac12 \sigma_{ik} \sigma^{ik} = \frac12 \left(
\sum_{i=1}^3 H_i^2 - 3H^2 \right) = \frac32 A H^2, \\
q &=& \frac{d}{d t} H^{-1}-1= -H^{-2} \left(\dot H + H^2\right).
\end{eqnarray}

The sign of the deceleration parameter indicates whether the model
inflates or not. The positive sign of $q$ corresponds to
``standard'' decelerating models whereas the negative sign
indicates inflation.

\section{Evolution of perfect cosmological fluids on Bianchi
type I brane Universes}
The line element of a Bianchi type I space-time, which generalizes
the flat Robertson-Walker metric to the anisotropic case, is given
by
\begin{equation}
ds^2 = - dt^2 + a_1^2(t) dx^2 + a_2^2(t) dy^2 + a_3^2(t) dz^2.
\end{equation}

We assume that the thermodynamic pressure $p$ of the cosmological
fluid obeys a linear barotropic equation of state $p=(\gamma-1)
\rho, \, \gamma=\text{const.}$ and $1 \leq \gamma \leq 2$.

Using the variables (\ref{V})-(\ref{H}) the gravitational field
equations and the Bianchi identity on the brane take the form
\begin{eqnarray}
3 \dot H + \sum_{i=1}^3 H_i^2 &=& \Lambda - \frac{3\gamma-2}2 \,
k_4^2 \rho - \frac{3\gamma-1}{12} \, k_5^4 \rho^2, \label{dH} \\
\frac1{V} \frac{d}{d t} (V H_i) &=& \Lambda - \frac{\gamma-2}2 \,
k_4^2 \rho - \frac{\gamma-1}{12} \, k_5^4 \rho^2, \quad i=1,2,3,
\label{dVHi} \\
\dot \rho + 3 \gamma H \rho &=& 0.
\end{eqnarray}

Thus, the time evolution of the energy density of the matter is
given by
\begin{equation}\label{rho}
\rho = \rho_0 V^{-\gamma}, \qquad \rho_0 = \text{const.} > 0.
\end{equation}

By summing Eqs. (\ref{dVHi}) we find
\begin{equation}\label{dVH}
\frac1{V} \frac{d}{dt} (VH) = \Lambda - \frac{\gamma-2}2 \, k_4^2
\rho - \frac{\gamma-1}{12} k_5^4 \rho^2.
\end{equation}
Subtracting Eq. (\ref{dVH}) back to Eqs. (\ref{dVHi}) we obtain
\begin{equation}\label{HiH}
H_i = H + \frac{K_i}{V}, \qquad i=1,2,3,
\end{equation}
with $K_i, \, i=1,2,3$ constants of integration satisfying the
consistency condition $\sum_{i=1}^3 K_i=0$. By using the evolution
law of the matter energy density the equation (\ref{dVH}), the
basic equation describing the dynamics of the anisotropic brane
world for a conformally flat bulk, can be written as
\begin{equation}
\ddot V = 3 \Lambda V - \frac{3(\gamma-2)}2 \, k_4^2 \rho_0
V^{1-\gamma} - \frac{\gamma-1}4 \, k_5^4 \rho_0^2 V^{1-2\gamma},
\end{equation}
and has the general solution
\begin{equation}\label{tt0}
t - t_0 = \int \left( 3\Lambda V^2 + 3 k_4^2 \rho_0 V^{2-\gamma}
+ \frac14 k_5^4 \rho_0^2 V^{2-2\gamma} + C \right)^{-1/2} d V,
\end{equation}
where $C$ is a constant of integration.

Therefore for a Bianchi type I induced brane geometry the general
solution of the gravitational field equations can be expressed in
the following exact parametric form, with $V\geq 0$ taken as
parameter:
\begin{eqnarray}
\theta &=& \left( 3\Lambda + 3 k_4^2 \rho_0 V^{-\gamma} + \frac14
k_5^4 \rho_0^2 V^{-2\gamma} + CV^{-2} \right)^{1/2},\label{theta}\\
a_i &=& a_{0i} V^{1/3} \exp \left[ K_i \int \left( 3\Lambda V^4 +
3 k_4^2 \rho_0 V^{4-\gamma} + \frac14 k_5^4 \rho_0^2 V^{4-2\gamma}
+ CV^2 \right)^{-1/2} d V \right], \quad i=1,2,3, \\
A &=& 3 K^2 \left( 3\Lambda V^2 + 3 k_4^2 \rho_0 V^{2-\gamma} +
\frac14 k_5^4 \rho_0^2 V^{2-2\gamma} + C \right)^{-1}, \\
\sigma^2 &=& \frac{K^2}2 V^{-2}, \\
q &=& 2 - \frac{36\Lambda V^2 + 18(2-\gamma) k_4^2 \rho_0
V^{2-\gamma} + 3(1-\gamma) k_5^4 \rho_0^2 V^{2-2\gamma}}{12\Lambda
V^2 + 12k_4^2 \rho_0 V^{2-\gamma} + k_5^4 \rho_0^2 V^{2-2\gamma} +
4C},
\end{eqnarray}
where $K^2=\sum_{i=1}^3 K_i^2$.

With the use of Eqs. (\ref{rho}), (\ref{HiH}) and (\ref{theta}),
from Eq. (\ref{dH}) it follows that the arbitrary integration
constants $K_i, \, i=1,2,3$ and $C$ must satisfy the consistency
condition
\begin{equation}
K^2=\frac23 C.
\end{equation}

At densities significantly greater than the nuclear one $\rho_n$,
e.g. $\rho>>\rho_n$, we have $p\to\rho$, with the speed of sound
$c_s$ tending to the speed of light, $c_s\to c$. A typical
approach to the nuclear equation of state in the very high-density
regime is to construct a relativistic Lagrangian that allows
``bare'' nucleons to interact attractively via scalar meson
exchange and repulsively via the exchange of a more massive vector
meson. But at the highest densities the vector meson exchange
dominates and one still has $p=\rho$. Therefore the equation of
state most appropriate to describe the high density regime of the
early Universe is the stiff Zeldovich one, with $\gamma=2$.

For a stiff cosmological fluid, the dynamics of the matter on the
brane is mainly determined by the correction terms from
$S_{\mu\nu}$, quadratic in the matter variables and due to the
form of the Gauss-Codazzi equations.

With $\gamma=2$ Eq. (\ref{tt0}) becomes
\begin{equation}
t - t_0 = \int \frac{V dV}{\sqrt{3\Lambda V^4 + \alpha V^2 +
\beta}},
\end{equation}
where $\alpha=3k_4^2\rho_0+C$ and $\beta=\frac14k_5^4\rho_0^2$.
The time dependence of the volume scale factor of the Bianchi
type I Universe in the high-energy limit is therefore given by
\begin{equation}
V(t) = \frac1{2\sqrt{3\Lambda}} e^{-\sqrt{3\Lambda}(t-t_0)}
\sqrt{\left( e^{2\sqrt{3\Lambda}(t-t_0)} - \alpha \right)^2 -
12\Lambda \beta}.
\end{equation}

For $t=t_S=t_0+\ln\left(2\sqrt{3\beta\Lambda}+\alpha\right)/
2\sqrt{3\Lambda}$ the volume scale factor is zero, $V(t_S)=0$. By
reparameterizing the initial value of the cosmological time
according to,
\begin{equation}
\exp(-2\sqrt{3\Lambda}t_0) = \alpha + 2 \sqrt{3\Lambda \beta},
\end{equation}
the evolution of the high-density brane Universe starts from
$t=0$ from a singular state $V_0=V(t=0)=0$.

The time evolution of the expansion, scale factors, anisotropy,
shear and deceleration parameter are given by
\begin{eqnarray}
\theta(t) &=& \sqrt{3\Lambda} \frac{e^{4\sqrt{3\Lambda}(t-t_0)} -
\alpha^2 + 12\Lambda\beta}{ \left( e^{2\sqrt{3\Lambda}(t-t_0)} -
\alpha \right)^2 - 12\Lambda\beta }, \label{2theta}\\
a_i(t) &=& a_{0i} e^{-\sqrt{\frac{\Lambda}3}(t-t_0)} \left[
\left( e^{2\sqrt{3\Lambda}(t-t_0)} - \alpha \right)^2 -
12\Lambda\beta \right]^{1/6} \exp \left[ 2K_i F \left(
e^{\sqrt{3\Lambda}(t-t_0)} \right) \right], \quad i=1,2,3, \\
A(t) &=& \frac{12K^2 e^{2\sqrt{3\Lambda}(t-t_0)} \left[ \left(
e^{2\sqrt{3\Lambda}(t-t_0)} - \alpha \right)^2 - 12\Lambda\beta
\right]}{ \left( e^{4\sqrt{3\Lambda}(t-t_0)} - \alpha^2 +
12\Lambda\beta \right)^2}, \\
\sigma^2(t) &=& \frac{ 6K^2 \Lambda e^{2\sqrt{3\Lambda}(t-t_0)}
}{ \left( e^{2\sqrt{3\Lambda}(t-t_0)} - \alpha \right)^2 -
12\Lambda\beta}, \\
q(t) &=& 12e^{2\sqrt{3\Lambda}(t-t_0)} \frac{ \alpha
e^{4\sqrt{3\Lambda}(t-t_0)} + (\alpha^2 - 12\Lambda\beta) \left(
\alpha - 2 e^{2\sqrt{3\Lambda}(t-t_0)} \right)}{ \left[
e^{4\sqrt{3\Lambda}(t-t_0)} - \alpha^2 + 12\Lambda \beta
\right]^2} - 1 \label{2q},
\end{eqnarray}
where $a_{0i}, \, i=1,2,3$ are arbitrary constants of integration
and $F(x)=\int \left[ (x^2-\alpha)^2 - 12\Lambda\beta
\right]^{-1/2} dx$.

If the cosmological constant is zero, $\Lambda=0$, the evolution
of the brane Universe in the extreme limit of high densities is
described by
\begin{eqnarray}
V(t) &=& \frac1{\sqrt{\alpha}} \sqrt{\alpha^2 (t-t_0)^2 -
\beta}, \\
\theta(t) &=& \frac{\alpha^2 (t-t_0)}{\alpha^2 (t-t_0)^2 -
\beta}, \\
a_i(t) &=& a_{0i} \left[ \alpha^2 (t-t_0)^2 - \beta \right]^{1/6}
\left[ \alpha(t-t_0) + \sqrt{\alpha^2 (t-t_0)^2 - \beta}
\right]^{K_i/\sqrt{\alpha}}, \quad i=1,2,3, \\
A(t) &=& \frac{3K^2}{\alpha^3} \frac{\alpha^2 (t-t_0)^2 -
\beta}{(t-t_0)^2}, \\
\sigma^2(t) &=& \frac{\alpha K^2}{2 \left[ \alpha^2 (t-t_0)^2 -
\beta \right]}, \\
q(t) &=& 2 + \frac{3\beta}{\alpha^2 (t-t_0)^2}.
\end{eqnarray}

\section{Scalar Field Evolution on the Anisotropic Bianchi
type I Brane}
In the previous Sections we have considered the evolution of
ordinary matter, obeying a barotropic equation of state, on the
brane-like Universe. In this case the gravitational field
equations can be solved exactly, with the general solution
represented in an exact parametric form for arbitrary $\gamma$
and in an exact analytic form in the extreme limit of high
densities, corresponding to $\gamma=2$.

As suggested by particle physics models, scalar fields, $\phi$,
with energy density and pressure given by $\rho_\phi =
\dot\phi^2/2 + U(\phi)$ and $p_\phi = \dot\phi^2/2 - U(\phi)$,
respectively, where $U(\phi)$ is the scalar field potential, are
supposed to play a fundamental role in the evolution of the early
Universe. For a Bianchi type I brane-Universe filled with a scalar
field the gravitational field equations take the form:
\begin{eqnarray}
3 \dot H + \sum_{i=1}^3 H_i^2 &=& k_4^2 \left[ U(\phi) -
\dot\phi^2 \right] - k_5^4 \left[ \frac16 \left(
\frac{\dot\phi^2}2 + U(\phi) \right)^2 + \frac14 \left(
\frac{\dot\phi^4}4 - U^2(\phi) \right) \right], \label{SdH} \\
\frac1{V} \frac{d}{d t}(VH_i) &=& k_4^2 U(\phi) - \frac1{12} k_5^4
\left( \frac{\dot\phi^4}4 - U^2(\phi) \right), \quad i=1,2,3.
\label{SdV}
\end{eqnarray}

The scalar field also obeys the following evolution equation:
\begin{equation}\label{ddphi}
\ddot \phi + 3 H \dot\phi + \frac{dU(\phi)}{d\phi} = 0.
\end{equation}
Depending on the functional form of the potential $U(\phi)$,
several cosmological scenarios of scalar field evolution on the
brane can be obtained.

As a first example we consider that the scalar field potential is
a constant, $U(\phi)=\Lambda=\text{const.}>0$. The potential acts
like a cosmological constant. Hence Eq. (\ref{ddphi}) gives
$\dot\phi=2\phi_0/V$, with $\phi_0>0$ a constant of integration.
Adding Eqs. (\ref{SdV}) we find the basic equation describing the
dynamics of the constant potential scalar field on the brane:
\begin{equation}\label{SddV}
\ddot V = \lambda V - \frac{\lambda_0}{V^3},
\end{equation}
where we denoted $\lambda=3k_4^2\Lambda+k_5^4\Lambda^2/4$ and
$\lambda_0=k_5^4\phi_0^4>0$.

The general solution of Eq. (\ref{SddV}) is given by
\begin{equation}
V(t) = \frac1{2\sqrt{\lambda}} e^{-\sqrt{\lambda}(t-t_0)} \sqrt{
\left( e^{2\sqrt{\lambda}(t-t_0)} - C \right)^2 - 4 \lambda_0
\lambda},
\end{equation}
with $C$ a constant of integration. This solution is similar to
the solution corresponding to the stiff matter cosmological fluid
and all the other physical quantities can be obtained from Eqs.
(\ref{2theta}-\ref{2q}) by means of the formal substitutions
$3\Lambda \to \lambda, \alpha \to C$ and $\beta \to \lambda_0$.
Therefore, by choosing the value of $t_0$ such that
$\exp(-2\sqrt{\lambda}t_0)=C+\sqrt{4\lambda_0\lambda}$, then the
brane Universe evolves from $t=0$ with a singular initial state.

As a second example of scalar field evolution on the brane we
consider a scalar field with potential energy proportional to the
kinetic one, so that $U(\phi)=(\gamma_\phi-1)\dot\phi^2/2,
\gamma_\phi=\text{const.}>1$ \cite{Ba1}. The energy density and
pressure of the scalar field are given by $\rho_\phi=\gamma_\phi
\dot\phi^2/2$ and $p_\phi=(2-\gamma_\phi)\dot\phi^2/2$,
respectively. The evolution equation for the scalar field gives
$\dot\phi=2\phi_0 V^{-1/\gamma_\phi}$. The evolution equation for
the volume scale factor becomes:
\begin{equation}
\ddot V = 6k_4^2 \phi_0^2 (\gamma_\phi-1) V^{1-2/\gamma_\phi} -
k_5^4 \phi_0^4 \gamma_\phi (2-\gamma_\phi) V^{1-4/\gamma_\phi},
\end{equation}
and has the general solution
\begin{equation}
t - t_0 = \int \left[ 6 k_4^2 \phi_0^2 \gamma_\phi
V^{2-2/\gamma_\phi} + k_5^4 \phi_0^4 \gamma_\phi^2
V^{2-4/\gamma_\phi} + C \right]^{-1/2} d V,
\end{equation}
with $t_0$ and $C$ arbitrary constants of integration. The time
variation of the physically important parameters is given in a
parametric form, with $V$ taken as parameter, by
\begin{eqnarray}
\theta &=& \sqrt{ 6k_4^2 \phi_0^2 \gamma_\phi V^{-2/\gamma_\phi}
+ k_5^4 \phi_0^4 \gamma_\phi^2 V^{-4/\gamma_\phi} + C V^{-2}},\\
a_i &=& a_{i0} V^{1/3} \exp \left[ K_i \int \left( 6k_4^2
\phi_0^2 \gamma_\phi V^{4-2/\gamma_\phi} + k_5^4 \phi_0^4
\gamma_\phi^2 V^{4-4/\gamma_\phi} + C V^2 \right)^{-1/2} d V
\right], \\
A &=& \frac{3K^2}{6k_4^2 \phi_0^2 \gamma_\phi V^{2-2/\gamma_\phi}
+ k_5^4\phi_0^4 \gamma_\phi^2 V^{2-4/\gamma_\phi} + C}, \\
\sigma^2 &=& \frac{K^2}2 V^{-2}, \\
q &=& 2 - \frac{ 18k_4^2 \phi_0^2 (\gamma_\phi-1)
V^{2-2/\gamma_\phi} + 3 k_5^4 \phi_0^4 \gamma_\phi (\gamma_\phi-2)
V^{2-4/\gamma_\phi} }{ 6k_4^2 \phi_0^2 \gamma_\phi
V^{2-2/\gamma_\phi} + k_5^4 \phi_0^4 \gamma_\phi^2
V^{2-4/\gamma_\phi} + C}, \\
\phi - \phi'_0 &=& 2\phi_0 \int \left[ 6k_4^2 \phi_0^2
\gamma_\phi V^{2-2/\gamma_\phi} + k_5^4 \phi_0^4 \gamma_\phi^2
V^{2-4/\gamma_\phi} + C \right]^{-1/2} V^{-1/\gamma_\phi} d V,
\end{eqnarray}
with $a_{i0}, i=1,2,3$ and $\phi'_0$ constants of integration. In
this model the scalar field satisfies an equation of state of the
form $p_\phi=(2/\gamma_\phi-1) \rho_\phi$.

The cosmological behavior of Universes filled with a scalar
field, $\phi$, as well as a Liouville type exponential potential
$U(\phi)=\Lambda e^{-k\phi}, \Lambda, k=\text{const.}>0$, has
been extensively investigated in the physical literature for both
homogeneous and inhomogeneous scalar fields (for a discussion of
the physical relevance of the exponential potential see
\cite{Co99} and references therein). The exponential type
potential arises in the four-dimensional effective Kaluza-Klein
type theories from compactification of the higher-dimensional
supergravity or superstring theories. In string or Kaluza-Klein
theories the moduli fields associated with the geometry of the
extra-dimensions may have effective exponential potentials due to
the curvature of the internal spaces or to the interaction of the
moduli with form fields on the internal spaces. Exponential
potentials can also arise due to non-perturbative effects such as
gaugino condensation.

For the flat isotropic Friedmann-Robertson-Walker geometry a
particular solution of the field equations with $\phi=\delta \ln
t$, has been obtained by Barrow \cite{Ba2}. In the framework of
general relativity there is no anisotropic Bianchi type I
solution of this type.

We try to generalize this solution for the case of the scalar
fields confined on the brane. By assuming that the constants $k$
and $\delta$ satisfy the condition $k \delta =2$, it is easy to
check by direct calculations that the following expressions
satisfy the field equations (\ref{SdH})-(\ref{ddphi}):
\begin{eqnarray}
V(t) &=& V_0 t^2, \qquad H(t) = \frac23 t^{-1}, \qquad a_i(t) =
a_{i0} t^{\frac23} \exp \left( -K_i V_0^{-1} t^{-1} \right),
\quad i=1,2,3, \\
A(t) &=& \frac34 K^2 V_0^{-2} t^{-2}, \qquad \sigma^2(t) = \frac12
K^2 V_0^{-2} t^{-4}, \qquad q(t) = \frac12, \\
\phi(t) &=& \frac2{\sqrt{3}k_4} \ln t, \qquad U(t) =
\frac2{3k_4^2} t^{-2}.
\end{eqnarray}

The scalar field and the potential are determined by the
four-dimensional coupling constant, with $\Lambda=2/3k_4^2,
k=\sqrt{3}k_4$ and $\delta=2/\sqrt{3}k_4$. The integrations
constants $K_i, i=1,2,3$ satisfy the consistency conditions
$\sum_{i=1}^3 K_i=0$.

However, if we substitute above results to the Eq. (\ref{SdH}),
then we will find a constraint on the constants of integration
$K^2=-8k_5^4V_0^2/27k_4^4$. From the expression of $K^2$ we can
see that for the general relativistic case, $k_5=0$, only
isotropic solution can exist, namely $K^2=\sum_{i=1}^3 K_i^2=0$,
leading to $K_i \equiv 0, \forall i$ and to a zero pressure scalar
field, $p_\phi \equiv 0$. For the brane world model, the
constraint condition for $K^2$ indicates that the scalar field
driven brane Universe with exponential type potential and with a
simple logarithmic time dependence of the scalar field cannot
exist, in general.

In the case of this simple model the scalar field potential can
also be given, in the isotropic case, as a function of the volume
scale factor, $U \sim V^{-1}$. It is interesting to find the
general solution of the field equations for this choice of $U$ for
the anisotropic brane, without imposing any supplementary
restrictions on the physical variables. For the sake of generality
we consider a more general problem, with a general functional
dependence of the scalar field potential on the volume scale
factor, $U=U(V)$. With this assumption the Klein-Gordon equation
for the scalar field (\ref{ddphi}) can be integrated to give
\begin{equation}
\dot\phi^2 = \frac2{V^2} \left( B - \int V^2 \frac{d U}{d V} d V
\right),
\end{equation}
and
\begin{equation}
\rho_\phi = \frac{\dot\phi^2}2 + U = \frac1{V^2} \left[ B + Y(V)
\right],
\end{equation}
where $B$ is a constant of integration and $Y(V)=2\int VU(V)dV$.
The general solution of the gravitational field equations on the
brane can be expressed again in an exact parametric form, with
$V$ as parameter:
\begin{eqnarray}
t - t_0 &=& \int \left\{ \frac32 K^2 + 3k_4^2 [B+Y(V)] +
\frac{k_5^4}4 V^{-2} [B+Y(V)]^2 \right\}^{-1/2} d V, \\
\theta &=& \sqrt{ \frac32 K^2 V^{-2} + 3k_4^2 V^{-2} [B+Y(V)]
+ \frac{k_5^4}4 V^{-4} [B+Y(V)]^2}, \\
a_i &=& a_{i0} V^{1/3} \exp \left\{ K_i \int \left[ \frac32 K^2
V^2 + 3k_4^2 V^2 [B+Y(V)] + \frac{k_5^4}4 [B+Y(V)]^2
\right]^{-1/2} d V \right\}, i=1,2,3, \\
A &=& \frac{3K^2}{\frac32 K^2 + 3k_4^2 [B+Y(V)] + \frac{k_5^4}4
V^{-2} [B+Y(V)]^2}, \\
\sigma^2 &=& \frac{K^2}{2V^2}, \\
q &=& 2 - \frac{18k_4^2 V Y'(V) + 3k_5^4 V^{-1} [B+Y(V)] Y'(V) -
3k_5^4 V^{-2} [B+Y(V)]^2}{6K^2 + 12k_4^2 [B+Y(V)] + k_5^4 V^{-2}
[B+Y(V)]^2}, \\
\phi - \phi_0 &=& \int \frac{\sqrt{2B + 2Y(V) - V Y'(V)} \, d V}
{\sqrt{\frac32 K^2 V^2 + 3k_4^2 [B+Y(V)] V^2 + \frac{k_5^4}4
[B+Y(V)]^2}},
\end{eqnarray}
with $t_0, a_{i0}, i=1,2,3$ and $\phi_0$ constants of integration
and $Y'(V)=d Y(V)/d V$.

Suppose that the scalar field potential is given, as a function of
the volume scale factor, by the simple relation $U=u_0/V$, with
$u_0>0$ a constant. Then, by denoting $\omega=3k_4^2u_0/2$ and
$\tau=(3K^2+2k_5^4u_0^2)/12k_4^2u_0$ and taking the integration
constant $B=0$, the general solution of the gravitational field
equations for the scalar field confined on the anisotropic brane
take the form:
\begin{eqnarray}
V(t) &=& \omega(t-t_0)^2 - \tau, \qquad \theta(t) =
\frac{2\omega(t-t_0)}{\omega(t-t_0)^2-\tau}, \\
a_{i}(t) &=& a_{i0} \prod_{j=1,2} \left[ \sqrt{\omega}(t-t_0) +
\varepsilon_j \sqrt{\tau}\right]^{1/3-\varepsilon_j K_i/2
\sqrt{\omega\tau}}, \quad i=1,2,3, \\
A(t) &=& \frac{3K^2}{4\omega^2(t-t_0)^2}, \qquad \sigma^2(t) =
\frac{K^2}{2[\omega(t-t_0)^2-\tau ]^2}, \\
q(t) &=& \frac12 + \frac{3\tau}{2\omega(t-t_0)^2}, \qquad \phi(t)=
\frac2{\sqrt{3}k_4}\sinh^{-1} \left[ \sqrt{\frac{\omega}{\tau}
(t-t_0)^2-1} \right], \\
U(t) &=& \frac{u_0}{\omega(t-t_0)^2-\tau}, \qquad
U(\phi)=\frac{u_0}{\tau} \sinh^{-2} \left( \frac{\sqrt{3}k_4}2
\phi \right),
\end{eqnarray}
where we have also taken $\phi_0=0$ and denoted
$\varepsilon_1=+1, \, \varepsilon_2=-1$.

\section{Perfect fluid Bianchi type V brane cosmologies}
The line element of a Bianchi type V space-time, representing the
anisotropic generalization of the open $k=-1$ Robertson-Walker
geometry, is given by
\begin{equation}
ds^2 = - dt^2 + a_1^2(t) dx^2 + a_2^2(t) e^{-2x} dy^2 + a_3^2(t)
e^{-2x} dz^2.
\end{equation}

The effective Einstein field equations on the brane are given by
\begin{eqnarray}
3 \dot H + \sum_{i=1}^3 H_i^2 &=& \Lambda - \frac{3\gamma-2}2 \,
k_4^2 \rho - \frac{3\gamma-1}{12} \, k_5^4 \rho^2, \label{VdH} \\
\frac1{V} \frac{d}{d t} (V H_i) -2 a_1^{-2} &=& \Lambda -
\frac{\gamma-2}2 \, k_4^2 \rho - \frac{\gamma-1}{12} \, k_5^4
\rho^2, \quad i=1,2,3, \label{VdV} \\
2 H_1 - H_2 - H_3 &=& 0, \label{HHH} \\
\dot \rho + 3 \gamma H \rho &=& 0.
\end{eqnarray}

From Eq. (\ref{HHH}) we obtain $a_2 a_3 = a_1^2$, leading to
$V=a_1^3$. The distribution of the matter energy-density on the
brane is given again by $\rho=\rho_0 V^{-\gamma}$. Adding Eqs.
(\ref{VdV}) we obtain
\begin{equation}\label{VdVH}
\frac1{V} \frac{d}{d t} (VH) = \Lambda + 2 V^{-2/3} -
\frac{\gamma-2}2 \, k_4^2 \rho - \frac{\gamma-1}{12} \, k_5^4
\rho^2.
\end{equation}
Subtracting Eq. (\ref{VdVH}) from Eqs. (\ref{VdV}) leads again to
\begin{equation}
H_i = H + \frac{K_i}{V}, \qquad i=1,2,3,
\end{equation}
with $K_i, \, i=1,2,3$ constants of integration satisfying the
consistency condition $\sum_{i=1}^3 K_i=0$.

Substitution of the Hubble parameters $H_i, \, i=1,2,3$ into Eq.
(\ref{HHH}) gives $2K_1=K_2+K_3$ and, consequently, we obtain
\begin{equation}
K_1 = 0, \qquad K_2 = - K_3.
\end{equation}

The general dynamics of the Bianchi type V space-time is
described by the equation
\begin{equation}
\ddot V = 6V^{1/3} + 3\Lambda V - \frac{3(\gamma-2)}2 \, k_4^2
\rho_0 V^{1-\gamma} - \frac{\gamma-1}4 \, k_5^4 \rho_0^2
V^{1-2\gamma},
\end{equation}
with the general solution
\begin{equation}
t - t_0 = \int \left( 9 V^{4/3} + 3\Lambda V^2 + 3k_4^2 \rho_0
V^{2-\gamma} + \frac14 k_5^4 \rho_0^2 V^{2-2\gamma} + C
\right)^{-1/2} d V,
\end{equation}
where $C$ is a constant of integration.

Therefore for a Bianchi type V induced brane geometry the general
solution of the gravitational field equations can be expressed in
the following exact parametric form, with $V\geq 0$ taken as
parameter:
\begin{eqnarray}
\theta &=& \left( 9V^{-2/3} + 3\Lambda + 3 k_4^2 \rho_0
V^{-\gamma} + \frac14 k_5^4 \rho_0^2 V^{-2\gamma} + CV^{-2}
\right)^{1/2}, \\
a_i &=& a_{0i} V^{1/3} \exp \left[ K_i \int \left( 9V^{10/3} +
3\Lambda V^4 + 3 k_4^2 \rho_0 V^{4-\gamma} + \frac14 k_5^4
\rho_0^2 V^{4-2\gamma} + CV^2 \right)^{-1/2} d V \right],
\quad i=1,2,3, \\
A &=& 3 K^2 \left( 9V^{4/3} + 3\Lambda V^2 + 3 k_4^2 \rho_0
V^{2-\gamma} + \frac14 k_5^4 \rho_0^2 V^{2-2\gamma} + C
\right)^{-1}, \\
\sigma^2 &=& \frac{K^2}2 V^{-2}, \\
q &=& 2 - \frac{ 72V^{4/3} + 36\Lambda V^2 + 18(2-\gamma) k_4^2
\rho_0 V^{2-\gamma} + 3(1-\gamma) k_5^4 \rho_0^2 V^{2-2\gamma} }{
36V^{4/3} + 12\Lambda V^2 + 12 k_4^2 \rho_0 V^{2-\gamma} + k_5^4
\rho_0^2 V^{2-2\gamma} + 4C},
\end{eqnarray}
where $K^2=\sum_{i=1}^3 K_i^2$.

Substitution of the Hubble parameter and of the density into Eq.
(\ref{VdH}) gives the second consistency condition satisfied by
the constants $K_i, i=2,3$:
\begin{equation}
K^2 = \frac23 C.
\end{equation}

Hence we can express the integration constants $K_2, K_3$ as
functions of the constant $C$:
\begin{equation}
K_2 = \pm \sqrt{\frac{C}3}, \qquad K_3 = \mp \sqrt{\frac{C}3}.
\end{equation}

In the limit of high matter densities ($\gamma=2$) the general
solution of the gravitational field equations for a Bianchi type
V geometry cannot be expressed in an exact analytic form.

\section{Discussions and final remarks}
In the present paper we have considered some classes of exact
solutions of the gravitational field equations on the brane world
in the framework of the Randall-Sundrum scenario. The
Randall-Sundrum mechanism was originally motivated as a possible
mechanism for evading Kaluza-Klein compactification by localizing
gravity in the presence of an uncompactified extra dimension. In
this model, the 5-dimensional bulk space-time is assumed to be
vacuum except for the presence of a cosmological constant. Matter
fields on the brane are regarded as responsible for the dynamics
of the brane. The quadratic terms in the energy-momentum tensor,
due to the form of the Gauss-Codazzi equations, may be important
in the very early stages of the Universe, leading to major changes
in the dynamics of the Universe.

For a perfect linear barotropic cosmological fluid confined on the
brane, the early cosmological evolution in anisotropic Bianchi
type I and V geometries is fundamentally changed by the inclusion
of the terms proportional to the squre of the energy density. The
time variation of the mean anisotropy parameter of the Bianchi
type I space-time is represented, for different values of
$\gamma$, in Fig.1. At high densities the brane Universe starts
its evolution from an isotropic state, with $A(t_0)=0$. The
anisotropy increases and reaches a maximum value after a finite
time interval $t_{\max}$, and for $t>t_{\max}$, the mean
anisotropy is a monotonically decreasing function, tending to zero
in the large time limit. This behavior is in sharp contrast to the
usual general relativistic evolution, illustrated by the case
$\gamma=1$ in Fig.1 (for this choice the quadratic contribution to
the energy-momentum tensor vanishes), in which the Universe is
born in a state of maximum anisotropy. In the initial stage the
evolution is non-inflationary, but in the large time limit the
brane Universe ends in an inflationary stage. In Fig.2 we present
the dynamics of the deceleration parameter for different values of
$\gamma$. In the limit of small $t$ (and small $V$, also), and
with $C=0$, from Eq. (\ref{tt0}) we obtain $V \sim t^{1/\gamma}$
(this result is valid for both Bianchi type I and V geometries).
Therefore in the early stages of evolution of the Universe the
mean anisotropy varies as $A \sim t^{2-2/\gamma}$. Near the
singular point the scale factors behave like $a_{i0}\sim
t^{1/3\gamma}\exp\left( \frac{\gamma}{1+\gamma} K'_i \,
t^{1+1/\gamma} \right)$. The deceleration parameter is given by
$q=3\gamma-1$. Figs.3 and 4 present the time evolution of the mean
anisotropy and deceleration parameter for the brane with Bianchi
type V geometry, the general features of the cosmological dynamics
being similar to the Bianchi type I case.

\begin{figure}
\epsfxsize=10cm \centerline{\epsffile{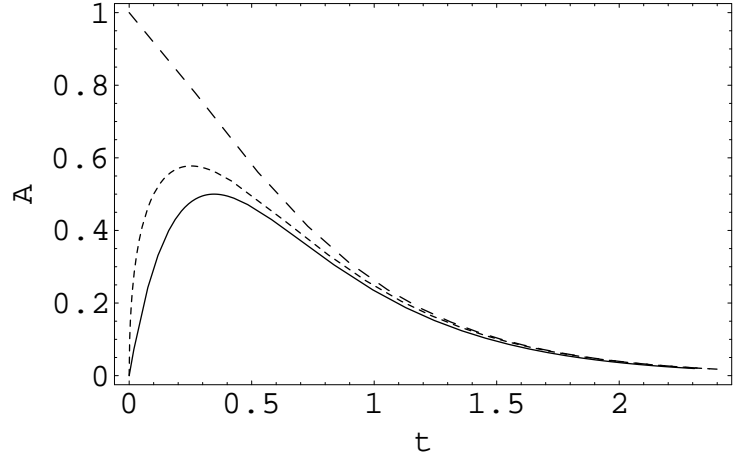}}
\caption{Variation as a function of time of the anisotropy
parameter of the Bianchi type I brane Universe with confined
perfect cosmological fluid for $\gamma=2$ (solid curve),
$\gamma=4/3$ (dotted curve) and $\gamma=1$ (dashed curve). The
constants have been normalized according to the conditions
$3\Lambda=1, 3k_4^2\rho_0=1, k_5^4\rho_0^2/4=1, C=1$ and
$K^2=2/3$.} \label{FIG1}
\end{figure}

\begin{figure}
\epsfxsize=10cm \centerline{\epsffile{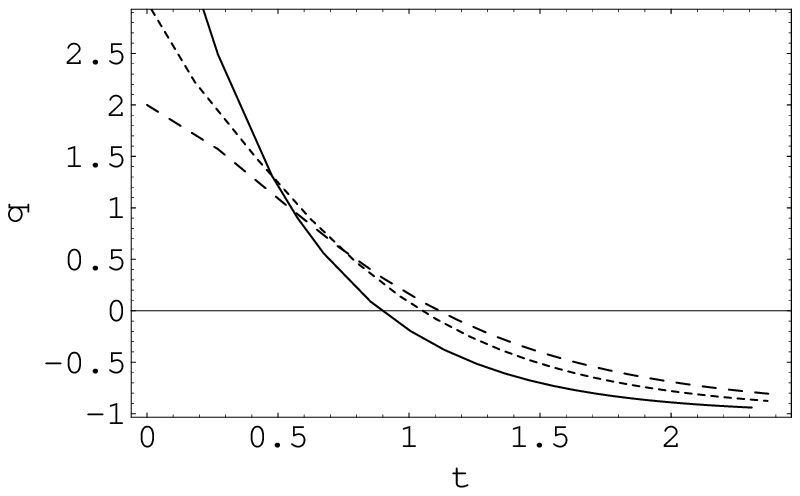}}
\caption{Variation as a function of the time of the deceleration
parameter $q$ of the Bianchi type I brane geometry, with perfect
cosmological fluid confined on the brane, for $\gamma=2$ (solid
curve), $\gamma=4/3$ (dotted curve) and for $\gamma=1$ (dashed
curve). The constants have been normalized according to the
conditions $3\Lambda=1, 3k_4^2\rho_0=1, k_5^4\rho_0^2/4=1$ and
$C=1$.} \label{FIG2}
\end{figure}

\begin{figure}
\epsfxsize=10cm \centerline{\epsffile{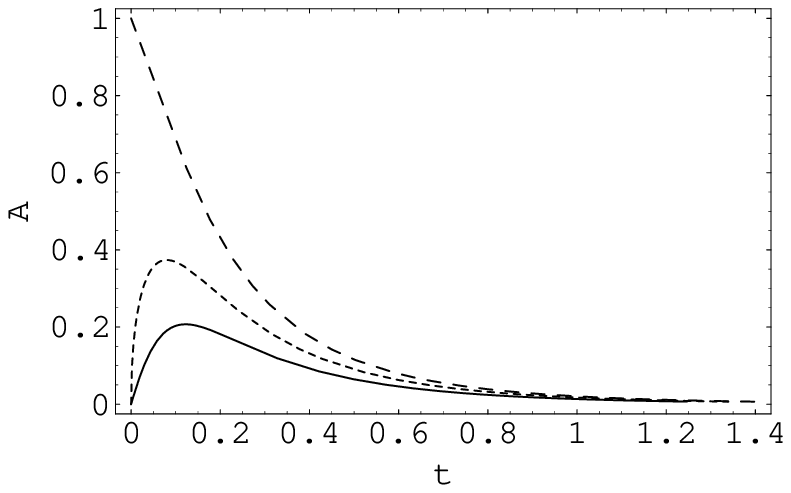}}
\caption{Variation as a function of time of the anisotropy
parameter of the Bianchi type V brane Universe with confined
perfect cosmological fluid for $\gamma=2$ (solid curve),
$\gamma=4/3$ (dotted curve) and $\gamma=1$ (dashed curve). The
constants have been normalized according to the conditions
$3\Lambda=1, 3k_4^2\rho_0=1, k_5^4\rho_0^2/4=1, C=1$ and
$K^2=2/3$.} \label{FIG3}
\end{figure}

\begin{figure}
\epsfxsize=10cm \centerline{\epsffile{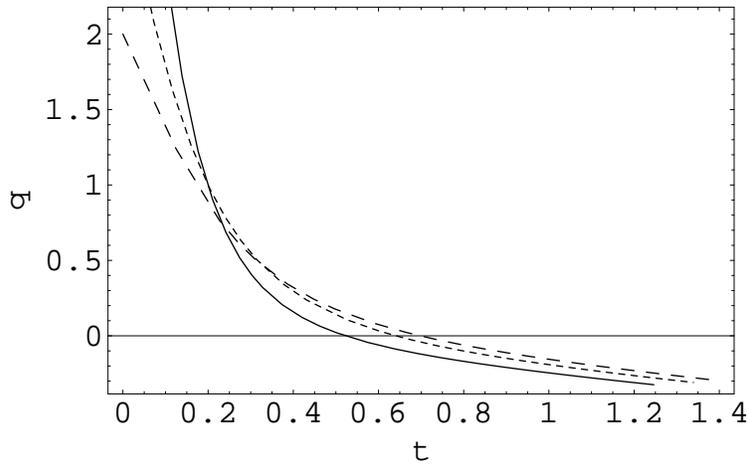}}
\caption{Variation as a function of the time of the deceleration
parameter $q$ of the Bianchi type V brane geometry, with perfect
cosmological fluid confined on the brane, for $\gamma=2$ (solid
curve), $\gamma=4/3$ (dotted curve) and for $\gamma=1$ (dashed
curve). The constants have been normalized according to the
conditions $3\Lambda=1, 3k_4^2\rho_0=1, k_5^4\rho_0^2/4=1$ and
$C=1$.} \label{FIG4}
\end{figure}

The general evolution of scalars fields confined on the
anisotropic brane is also modified by the presence of the
quadratic terms in the energy-momentum tensor. For a constant
potential scalar field, the solution of the gravitational field
equations is very similar to the stiff cosmological fluid case,
with the cosmic evolution starting from an isotropic state, with
zero anisotropy and with a maximum anisotropic phase reached after
a finite time $t_{\max}$. The dynamics of the scalar field
satisfying the equation of state
$p_\phi=(2/\gamma_\phi-1)\rho_\phi$, $\gamma_\phi>1$, depends on
the values of $\gamma_\phi$. For $1<\gamma_\phi<3/2$, the Universe
is born in an isotropic state, but for $\gamma_\phi>3/2$ the
behavior is similar to the general relativistic case, with an
infinite initial anisotropy, which rapidly decays to zero. For
small $\gamma_\phi$ the evolution is non-inflationary. In the
limit of large $\gamma_\phi$ the scalar field behaves like a
cosmological fluid obeying an equation of state of the form
$p_\phi=-\rho_\phi$. In this case and for large times the behavior
of the Universe is described by
\begin{eqnarray}
V &=& V_0 \exp(H_0 t), \qquad H = \frac13 H_0, \qquad a_i =
a_{i0} \exp \left( \frac{H_0}3 t - \frac{K_i}{V_0 H_0} e^{-H_0 t}
\right), \quad i=1,2,3, \\
A &=& \frac{3K^2}{V_0^2 H_0^2} \exp(-2H_0 t), \qquad q = -1,
\qquad \phi = 2\phi_0 t, \qquad \rho_\phi = 2 \phi_0^2 \gamma_\phi
= \text{const.},
\end{eqnarray}
where $H_0=\phi_0 \sqrt{6k_4^2 \gamma_\phi + k_5^4\phi_0^2
\gamma_\phi^2}$. In the large time limit the Universe ends in an
isotropic de Sitter inflationary phase. The time variations of the
anisotropy and deceleration parameters are represented, for
different values of $\gamma_\phi$, in Figs. 5 and 6.

\begin{figure}
\epsfxsize=10cm \centerline{\epsffile{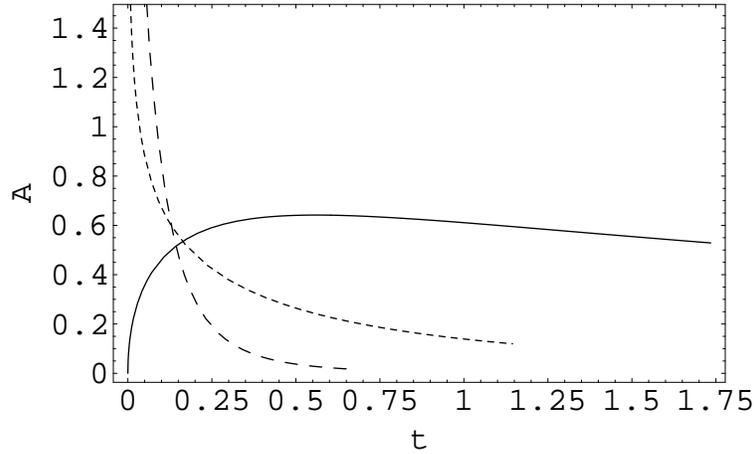}}
\caption{Variation as a function of time of the anisotropy
parameter of the Bianchi type I brane Universe with confined
scalar fields obeying the equation of state
$p_\phi=(2/\gamma_\phi-1) \rho_\phi$ for $\gamma_\phi=1.5$ (solid
curve), $\gamma_\phi=2$ (dotted curve) and $\gamma_\phi=5$ (dashed
curve). The constants have been normalized according to the
conditions $6k_4^2\phi_0^2=1, k_5^4\phi_0^4=1, C=1$ and $K^2=1$.}
\label{FIG5}
\end{figure}

\begin{figure}
\epsfxsize=10cm \centerline{\epsffile{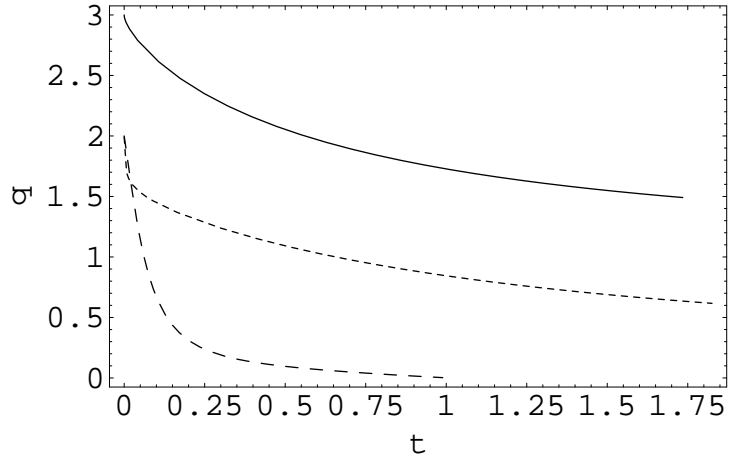}}
\caption{Variation as a function of time of the deceleration
parameter $q$ of the Bianchi type I brane Universe with confined
scalar fields obeying the equation of state
$p_\phi=(2/\gamma_\phi-1) \rho_\phi$ for $\gamma_\phi=1.5$ (solid
curve), $\gamma_\phi=2$ (dotted curve) and $\gamma_\phi=5$ (dashed
curve). The constants have been normalized according to the
conditions $6k_4^2\phi_0^2=1, k_5^4\phi_0^4=1$ and $C=1$.}
\label{FIG6}
\end{figure}

Scalar fields with potentials of the form $U(\phi)=U_0 \sinh^\beta
(\alpha\phi), \, U_0, \beta, \alpha=\text{consts.}$, have been
recently considered in quintessence models as tracker solutions of
the gravitational field equations, which can drive the Universe in
its present non-decelerating phase \cite{UrMa00}. For such a type
of potential we have obtained the general solution of the field
equations in the case of the anisotropic brane world for $\beta
=-2.$ The asymptotic behavior of the potential corresponds to an
inverse power-law like form for small $\phi$, $U(\phi)\sim
\phi^{-2}$ and to an exponential one for large values of the
scalar field, $U(\phi)\sim e^{-\sqrt{3}k_4\phi}$. In this model
the Universe starts from a state with infinite anisotropy and ends
in an isotropic phase, but the evolution is generally
non-inflationary, with $q>0$ for all times.

It is interesting to note that in two measure gravitational
theories with scale invariance a dilaton field with exponential
type potentials has to be also introduced, leading to non-trivial
mass generation. Interpolating models for natural transition from
inflation to a slowly accelerated Universe at late times appear
naturally \cite{Gu200}. For cosmological models without the
cosmological constant problem and theories of extended objects
(strings, branes) without a fundamental scale see Guendelman
\cite{Gu300,Gu400}.

We have not presented the solutions corresponding to scalar fields
confined to the Bianchi type V brane world. Their mathematical
form is very similar to the Bianchi type I case, with one
correction term, proportional to $V^{4/3}$ added to the parametric
time equation. Due to the presence of this term, exact analytic
solutions cannot be generally obtained for scalar field models in
Bianchi type V space-times, but the general physical behavior of
scalar fields in both geometries has the same qualitative
features.

\section*{Acknowledgments}
The work of CMC is supported by the Taiwan CosPA project and, in
part, by the Center of Theoretical Physics at NTU and National
Center for Theoretical Science.

\end{document}